\documentclass[prc,showpacs,showkeys,superscriptaddress,nofootinbib,floatfix,onecolumn]{revtex4}
\usepackage[cp1251]{inputenc}
\usepackage[T2A]{fontenc}
\usepackage[english]{babel}
\usepackage{color}
\usepackage{amsfonts}
\usepackage{amsbsy}
\usepackage{amssymb}
\usepackage{mathrsfs}
\usepackage{graphicx}
\usepackage{subfigure}
\newcommand{\bea}{\begin{eqnarray}}
	\newcommand{\eea}{\end{eqnarray}}
\begin{document}
	\title{Sign problems in path integral formulations of quantum mechanics \\ and quantum statistics}
	
	\author{Vladimir Filinov, Alexander Larkin}
	
	\address{Joint Institute for High Temperatures of the Russian Academy of Sciences, \\Izhorskaya 13 Bldg 2, Moscow 125412, Russia
	}
	
	
	\begin{abstract}
		Nowadays the term 'sign problem' is used to identify two different problems. 
		
		The ideas to overcome the first type of the 'sign problem' of strongly oscillating complex valued imtegrand  in the Feynman path 
		integrals comes from Picard-Lefschetz theory and a complex version of Morse theory. 
		The main idea is to select Lefschetz thimbles as the cycle 
		approaching the critical point at the path-integration, where the imaginary part of the complex action  
		stays constant. Since the imaginary part of the action is constant on each thimble, 
		the sign problem disappears and the integral can be calculated much more effectively. 
		Here based on the Metropolis -- Hastings algorithm a new method of calculations of the integral of the strongly oscillating 
		integrands has been prosed. 
		Some simple test calculation and comparison with available analytical results have been carried out. 
		
		The second type of the 'sign problem' arises at studies Fermi systems by path integral approach and is 
		caused  by the requirement of antisymmetrization of the real valued matrix elements of the density matrix. 
		An explicit analytical expression for effective pair pseudopotential 
		in phase space has been discussed in Wigner formulation of quantum mechanics. 
		Obtained pseudopotential allow to account for Fermi statistical effects as realizes 
		the Pauli blocking of fermions due to the repulsion between 
		identical fermions, which prevents their occupation of same phase space cell.  
		To test this approach, calculations of  the momentum distribution function of the 
		ideal system of Fermi particles have been presented over a wide range of 
		momentum and degeneracy parameter.  
		
	\end{abstract}
	
	
	\maketitle
	
	\section{Introduction}

One of the major difficulty for the Path Integral Monte Carlo (PIMC) simulation 
of the quantum systems 
 of particles is so called 'sign problem'. 
 The major  aim of this paper is to discussed the new effective Monte Carlo methods for numerical simulations 
 of path integrals with a sign problems as these methods can also provide a new perspectives  
 in path integration. Our aim is to consider techniques of universal character, providing 
 insights to this problems. However now a days the term 'sign problem' is used to identify two different 
 problems.  
 
 The first type of the 'sign problem' arises in the  Wigner and Feynman path integral formulation of 
 quantum mechanics and the finite density quantum chromodynamics, where a complex 
 action does not give a real and positive Boltzmann -- like weight (for example, by the by the Wick rotation) 
 to resort to the traditional Monte Carlo methods. 
The so-called reweighting algorithm  is highly ineffective when the imaginary part 
 of the action becomes very large, because one needs to take a sample from a configuration space, where the 
 weights of nearby configurations have almost the same amplitudes but very different phases. 
 
 There has been many proposals to circumvent the 'sign problem'. 
Basic possible approach to this problem is to consider the variables, which are 
 assumed to be real in the original formulation, to be complex and to extend the cycle of
 path-integration to a complex space in order to achieve better convergence. 

So long under the typical physical conditions 
the integrand is holomorphic in the new complex variables and the final value 
of the path integral is unchanged by Cauchy▓s theorem. Making use of 
the Picard-Lefschetz theory and  a complex version of Morse theory we can select the cycle approaching 
the saddle point at the path-integration, where the imaginary part of the complex action  stays constant 
 (Lefschetz thimbles) \cite{fedoryuk1976asymptotics,witten2011analytic}. 
 Since the imaginary part of the action is constant on each thimble, 
 the sign problem disappears and the integral can be calculated much more effectively. 
 
 However, since different thimbles are strongly separated, one needs to develop  
 method to incorporate contributions from all relevant thimbles.
 To overcome this problem  a natural choice is to use a gradient flow of action (described below), 
 starting from the original real space of integration and creating a new manifold at a finite flow time, 
which is equivalent to the integration over original real space  
 \cite{alexandru2016monte,alexandru2017schwinger}. 
In particular, when the flow time  approaches infinity, the new manifold 
is composed of Lefschetz thimbles. 
In practise, as long as the finite flow-time is large enough 
then the 'sign problem' will be alleviated, as integral turn into integrals of an oscillating 
function with decaying amplitude. However reducing the amount of flow time may 
also reduce the effectiveness against the 'sign problem' and the multimodal problem simultaneously.
 
The alternative approach of sampling at calculation of the path-integrals on thimbles 
is based on the making use of the complexified Langevin equation 
\cite{parisi1988complex,klauder1983langevin,klauder1984coherent}. 
Of course, application of the noise  leads to departures from the thimble, that accumulate and need 
to be corrected. 
Numerically, this procedure can be made stable. 

Besides, the Langevin algorithm, other algorithms have been proposed: 
an Hybrid Monte Carlo algorithm \cite{fujii2013hybrid}, that however  is essentially 
as expensive as the Langevin algorithm; two Metropolis algorithms 
\cite{mukherjee2013metropolis,alexandru2016monte} that are simpler and faster, but have
the risk of poor acceptance for large systems and an alternative algorithm \cite{di2015thimble} 
that ensures a control of the thimble at the price of limited scalability. 

In this work we present the new Metropolis -- Hastings algorithm for searching critical points and 
subsequent generation of the a Lefschetz thimbles.  The algorithm allows to find critical points and to sample 
initial conditions for downward flows from the different quarters or halfs of vicinity 
of the critical point. To increase efficiency of numerical procedure we have separated 
the Markovian transition on the complex plane in two sub-steps: the proposal and the acceptance-rejection. 
The proposal probability allow  to propose a new state for given one.  Here we are free in choosing 
proposing probability as it affect only the efficiency of the sampling the main contribution to the 
integrals and does not change the final result of calculations. The acceptance distribution is the 
conditional probability to accept/reject the proposed state. 

The integrals involved in our test calculations are one variable integrals and can be performed 
analytically or independent numerically. However, it provides an interesting benchmark which can be
seen as a limiting case of more realistic path integrals. It is non-trivial from the point of view 
of a Monte Carlo integration. For comparison with analytical results we present 
some calculations for the Airy function  and some results 
on the Fourier transform of basic 1D factors comprising the Wiener path integral  
representation of the Wigner function in phase space. 
It also provides a case where different aspects of our approach can be clearly visualized.

 The second type of the 'sign problem' arises from the requirement of antisymmetrization of the 
 matrix elements of the fermion 
 density matrix. A a result all thermodynamic quantities are presented as the sum of real valued terms 
 with alternating sign related to even and odd permutations and are equal to the small difference of two 
 large numbers, which are the sums of positive and negative terms. The numerical calculation in this case is 
 severely hampered. To overcome this issue a lot of approaches have been developed but  
 the 'sign problem' for strongly correlated 
 fermions has not been fully solved during the last fifty years. 
 Let us mention new original approaches developed in \cite{EbelForFil,njp,prl2,cpp,prl1}. 
 
 Monte Carlo simulations at finite temperature over the entire fermion densities
 range down to half the Fermi temperature have been carried out by permutation
 blocking path integral Monte Carlo (PB-PIMC) approach \cite{njp,prl2}.
 For purpose to simulate fermions in the canonical ensemble, it was
 combined a fourth-order approximation of density matrix derived with
 a full antisymmetrization on all time
 slices in discrete versions of the paths. It was
 demonstrated that this approach effectively allows for the combination
 of $N!$ configurations from usual path integral MC into 
 a single configuration weight of PB-PIMC, thereby reducing the complexity of
 the problem.
 Treatment of interacting fermions  in \cite{prl1}
 has been carried out at very high densities. Obtained results for finite number of 
 particles were extrapolated to the thermodynamic limit \cite{prl2}.  
 
 The configuration path integral Monte Carlo (CPIMC) approach \cite{cpp,prl1}
 for degenerate correlated fermions with arbitrary pair interactions
 at finite temperatures is based on representation of the
 N-particle density operator in a basis of (anti-)symmetrized
 N-particle states (configurations of occupation numbers).
 The main idea of this approach is to evaluate the path integral
 in space of occupation numbers instead of configuration space. 
 This leads to path integrals occupation
 number representation allowing to treat arbitrary pair interactions in 
 a continuous space. However it turns out that 
 CPIMC method exhibits a complementary behavior and works well at 
 weak nonideality and strong degeneracy. Unfortunately, the
 physically most interesting region, where both fermionic exchange
 and interactions are strong simultaneously remains out of reach.
 
 Another approach was proposed in \cite{larkin2017pauli}, where 
 to avoid the 'sign problem' and to realize Pauli blocking of fermions 
 accounting for the Fermi statistical effects without antisymmetrization of the matrix elements 
 the Wigner formulation of quantum mechanics in phase space has been used. 
 In \cite{larkin2017pauli,larkin2017peculiarities} it  has been shown that accounting for only the identical and 
 pair permutations allow to derive an effective pair pseudopotential in phase space  
 realizing the Pauli blocking. The derived repulsive pseudopotential depends on coordinates, momenta, 
 and the degeneracy parameter of fermions  and do not allow  two identical fermions to  
 to occupy the same cell in phase cell. A new quantum phase space path integral Monte Carlo 
 method has been also developed to calculate average values of arbitrary quantum operators in phase space.  
 In this work we briefly present the basic ideas of this approach and test results for ideal degenerated fermions. 
Results for strongly interacting fermions in wide region of temperature and density for electron -- hole plasma 
are pesented in \cite{EbelForFil}. 
\section{The 'sign problem' of the complex valued integrand in path integrals.}
To explain the basic ideas we consider a one-dimensional quantum-mechanical system consisting of one particle in a potential field $ U (q) $.
The Hamilton function of such a system has the form:
\begin{equation}
	\label{eq_FP_hamiltonian}
	H(p,q) = \frac{p^2}{2m} + U(q),
\end{equation}
where $ p $ and $ q $ are the momentum and coordinate of the particle correspondingly.
The potential function $ U (q) $ is attractive and finite when $q$ is real.
Also we assume that the system is in a state of thermodynamic equilibrium with some external thermostat.
In other words, we consider a canonical ensemble with temperature $ T $ and volume $ V $ (one-dimensional).

The quantum canonical ensemble can be fully characterized by the Wigner function $ W (p, q) $, which is 
essentially a density matrix in the mixed coordinate-momentum representation and is defined 
as the Fourier transform of the density matrix in the coordinate representation: 
\begin{equation}
	\label{eq_FP_wignerfunction}
	W(p,q) = \int\limits_{-\infty}^{+\infty} {\rm{d}} \xi \, {\rm{e}}^{\frac{{\rm{i}}}{\hbar} p \xi} \, \langle q - \xi/2 | \, {e}^{ -\beta \hat{H} } \, | q + \xi/2 \rangle,	
\end{equation} 
where $ \beta = \frac {1} {kT} $ is the value inversely proportional to temperature, 
$ \hat {H} $ is the Hamilton operator obtained from (\ref{eq_FP_hamiltonian}) by replacing 
$ p $, $ q $ on the momentum and the coordinates operators $\hat{p} $, $ \hat{q} $ respectively. 
The Wigner function $ W (p, q) $ can be formally considered as a generalization of the classical 
distribution function in phase space to quantum mechanical systems.
At the same time, this generalization should not be taken literally, because, although the Wigner 
function is real, it can take both positive and negative values.

As example, let us consider the momentum distribution function for the canonical ensemble, which can be obtained from 
\begin{equation}
	\label{eq_FP_distrfpdensmatr}
	W(p) = \int\limits_{-\infty}^{+\infty} {\rm{d}}q \int\limits_{-\infty}^{+\infty} {\rm{d}}\xi \, 
	{\rm{e}}^{\frac{{\rm{i}}}{\hbar} p \xi}  \, \langle q - \xi/2 | \, {\rm{e}}^{ -\beta \hat{H} } \, | q + \xi/2 \rangle.
\end{equation}
\section{Path integral representation}  
Now we need to calculate the density matrix (the integral in (\ref {eq_FP_distrfpdensmatr})). 
In the general case, this cannot be done directly, since the kinetic and potential energy operators 
entering the Hamiltonian are non-commutative, and the statistical operator $ \exp (- \beta \hat{H}) $ 
does not split into the product of the operators 
$ \exp (- \frac {\beta  hat{p} ^ 2} {2m}) $ and $ \exp (- \beta U (\hat {q})) $.
Therefore, we use the following procedure, leading to the representation of the density matrix 
in the form of a path integral.
We break the statistical operator $ \exp(- \beta\hat{H}) $ into the product $ 2K $ of the 
same operators $ \exp(- \varepsilon \hat{H}) $, where 
$ \varepsilon = \frac{\beta} {2K} $, assuming the integer $ K $ is large enough.
Then between these operators we insert $ 2K-1 $ of the unit operators 
$ \hat{1} $, which we respectively replace by completeness relations for states
 with a certain coordinate $ | q \rangle $: 
\begin{eqnarray}
	\label{eq_PI_densmatr_inmd1}
	\nonumber
\langle q-\xi/2 | \, {\rm{e}}^{-\beta \hat{H}} \, | q + \xi/2 \rangle
	\, {e}^{-\varepsilon \hat{H}} \, \ldots {e}^{-\varepsilon \hat{H}} \, | q + \xi/2 \rangle = 
&&\\ \nonumber
= \langle q-\xi/2 | \, {\rm{e}}^{-\frac{\beta}{2 K} \hat{H}} \, \hat{1} \, \ldots \, \hat{1} \, {\rm{e}}^{-\frac{\beta}{2 K} \hat{H}} \, | q + \xi/2 \rangle
&& \\ 
= \int \ldots \int {\rm{d}}q_{-K+1} \ldots {\rm{d}}q_{K-1} \,
\Biggl[ \prod\limits_{k = -K}^{K-1} \langle q_{k+1} | \, 
{\rm{e}}^{-\frac{\beta}{2 K} \hat{H}} \, | q_{k} \rangle \Biggr].
\end{eqnarray} 

For a small $ \varepsilon $ corresponding to a large $ K $, 'high-temperature' statistical operators 
$ \exp (- \varepsilon \hat{H}) $ can be decomposed into the product of the operators 
$ \exp(- \frac{\varepsilon \hat{p} ^ 2} {2m}) $ and $ \exp(- \frac{\varepsilon}{2}U(\hat{q})) $ up to $ O (\varepsilon^2) $ .
After that, the corresponding matrix elements can be calculated directly using the completeness relations 
for states with a certain momentum $ | p \rangle $, plane waves $ \langle q | p \rangle = \frac{{i}} {\hbar} p q $ 
and taking into account the fact that the states $ | p \rangle $ and $ | q \rangle $ are proper for the operators $\hat {p} $ and $ \hat {q} $ 
respectively. A small parameter $ {\Delta \tau} = \varepsilon \hbar $ has the dimension of time. 
We substitute the 'high-temperature' matrix elements 
into the formula (\ref{eq_PI_densmatr_inmd1}) and  
pass from the integration variables $ q $ and $ \xi $ to the variables $ q_{- K} = q + \xi /2 $ and 
$ q_{+ K} = q - \xi /2 $. As a result, we obtain the path integral representation of the Wigner function. 

For example, the momentum distribution function can be written 
in the form of a multiple integral over the coordinates $ q_{k} $, $ k = 0, \, \pm 1, \, \ldots, \, \pm K $:
\begin{eqnarray}
	\label{eq_PI_distrfpdiscr}
W(p,q) = \left(\frac{2 \pi \hbar {\Delta\tau}}{m} \right)^{-K} \, \int {\rm{d}}{\xi} \int \dots \int {\rm{d}}q_{-K+1} \ldots {\rm{d}}q_{K-1} {\rm{e}}^{-\Phi_{2K+1}(q_{-K}, \ldots, q_{K}) } + O({\Delta\tau}),
&& \\ \nonumber
\Phi_{2K+1}(q_{-K}, \ldots, q_{K}) =-\frac{{\rm{i}}}{\hbar}p \xi + \frac{{\Delta\tau}}
{\hbar} \sum\limits_{k =-K}^{K-1} \left[\frac{m}{2} \left(\frac{q_{k+1}-q_{k}}{{\Delta\tau}}
\right)^2
+ \frac{U(q_{k+1}) + U(q_{k})}{2} \right]. 
\end{eqnarray}
%
The parameter $ {\Delta \tau} = \frac {\beta \hbar} {2K} $ tends to zero for $ K \to \infty $, while the formula becomes accurate.
However calculation of the integrals like (\ref{eq_PI_distrfpdiscr}) is practically impossible due to the 
mentioned above 'sign problem' related to the strong oscillations  of the complex valued integrand. 

For  power dependence potential field the discrete expression (\ref{eq_PI_distrfpdiscr}) consists of the following factors: 
\begin{eqnarray}
	\label{eq_AS_intikreal}
	I_{k} &=& \int\limits_{-\infty}^{+\infty} {d}q_{k} \, \exp\left\{ a_{k} q_{k}^2 + b_{k} q_{k} + c_{k} + d_{k} q_{k}^{\alpha} \right\} 
\end{eqnarray}
where integration is carried out over the real axis of $ q_{k}  \in C_{\mathbb{R}} $

The integrand is also be analytic, which allow us to use the Cauchy's integral theorem and deform 
the integration contour on the complex plane $ q_{k} \in \mathbb{C} $ with the constant value of the integral $ I_{k} $.
(Note that the analytical structure of the integrand is mainly determined by the function $ q^{\alpha} $, 
since the polynomial $ a_{k} q_{k}^2 + b_{k} q_{k} + c_{k} $ is , obviously, a analytic single -- valued function.) 
Let us stress that detailed consideration of the (\ref{eq_AS_intikreal})--like integrals 
for nonanalytical action is presented in \cite{fedoryuk1989asymptotic}. 

\section{Basics of the complexification on Lefschetz thimbles}\label{semi:model}

According to the Morse theory 
\cite{vladimir1988singularities,fedoryuk1976asymptotics,pham1983vanishing,berry1990hyperasymptotics,howls1997hyperasymptotics,witten2011analytic}  
the regions of integration over each real $q_k$  
in the path integrals (\ref{eq_PI_distrfpdiscr}), (\ref{eq_AS_intikreal}) 
is equivalent to the integration over complex valued set of Lefschetz thimbles which is homologically 
equivalent due to the Cauchy's integral theorem to the integration on real cycle  $C_{\mathbb{R}}$. 
Assuming  that $q_k$  takes the complex values $q_k \in \mathbb{C} $ and 
the action $\Phi[q_k(\tau)]$ is extended to a holomorphic function of $q$ 
let us condider the set $\Sigma$  of critical points (sadle points) 
$q_{k,\sigma}$,  which satisfy condition 
$\partial \bar{\Phi}[\bar{q}]/\partial\bar{q}|_{\bar{q}=\bar{q}_{k,\sigma}} =0$. 
The real Morse function in our case can be defined as 
$h \equiv - \mathbb{\Re}  \{\Phi[q_k]\} $ and  the associate gradient (downward) flow equations is given by: 
\begin{equation}
\label{difur}
\frac{ {\rm{d}} \mathbf{q} }{ {\rm{d}} l } =-\overline{ \frac{\partial {\Phi}(\mathbf{q})}{\partial \mathbf{q}} }, \qquad l \in \mathbb{R}.
\end{equation}
The Morse function $h$ is always strictly decreasing along a flow. 
Associated with a critical point $q_{k,\sigma}$, a Lefschetz thimble $\Upsilon_{\sigma}$ 
\cite{cristoforetti2012new} is defined 
by the union of all downward flows,  which trace back to $q_{k,\sigma} $ at $l \to - \infty$. 
Let us note that if $q_k(l)$ equals a critical point at some $l$, 
then the flow equation implies that $q_k(l)$ is constant for all $l$. So a non constant flow can 
only reach a critical point at $l \to - \infty$. 

One can also introduce another submanifold $\Pi_{\tau}$ of 
by the union of  all upward flows, satisfying equation with opposite sign to equation (\ref{difur})
\begin{equation}
\label{difur1}
\frac{d}{dl}\tilde{q}_k(l) = \frac{\partial \bar{\Phi}[\bar{\tilde{q}}_k]}{\partial\bar{\tilde{q}}_k}  , \, l \in \mathbb{R}^1 
\end{equation}
which converge to $q_{k,\tau}$ at  $l \to - \infty$, so that its intersection number with $q_k(l) \in \Upsilon_{\sigma}$ 
is unity and vanishing otherwise, $< \Upsilon_{\sigma} , \Pi_{\tau}  > = \delta_{\sigma,\tau}$. 

Strictly speaking  this means that $q_k(l)\in \Upsilon_{\sigma}$ if $q_k(l)$  is the solution of Eq.~(\ref{difur}) 
and for any positive (may be very small) $\epsilon >0$ there exists the positive  
(may be very large) $L>0$ that for all negative $l$ such that $-l>L$ we have  $\vert| q_k(l)-q_{k,\sigma} \vert| <\epsilon$.     
This allow in numerical simulation to use the following approximation. 
Due to the restrictions on computational time at solving equation (\ref{difur}),  (\ref{difur1}) it is necessary to 
exclude the small $\epsilon$ vicinity of critical point, where the parameter $l \to - \infty$. 
So in numerical simulations the staring points for downward flow have to be chosen  out side of a small vicinity  of $q_{k,\sigma}$  
and the averaging results of calculations by the Monte Carlo method can be done over ensemble of the  downward flows related to 
decreasing small $\epsilon$.  
Algorithm of this MC approach, sampling the  main contribution in integrals like (\ref{eq_PI_distrfpdiscr}), 
(\ref{eq_AS_intikreal})  will be discussed below.  

Then, according to Morse theory \cite{cristoforetti2012new}, it follows that  
\begin{equation}
\label{hh}
C_{\mathbb{R}} = \sum_{\sigma \in \Sigma} n_{\sigma} \Upsilon_{\sigma}, \qquad n_{\sigma}=<C_{\mathbb{R}}|\Pi_{\tau} >
\end{equation}
which holds in the homological sense. 
Now, for example, the momentum distribution function  Eq.~(\ref{eq_FP_distrfpdensmatr}) can be given by the formula 
\begin{eqnarray}
\label{FinFp} 
W(p) &=& \sum_{\sigma \in \Sigma} n_{\sigma } {e}^{ -\rm{i} \mathbb{\Im} (\Phi[q(\tau)]) } 
\int_{\Upsilon_{\sigma}} {D}q(\tau) \, {e}^{ -\mathbb{\Re} (\Phi[q(\tau)] )}, \, .
\end{eqnarray} 
As consequence, for the critical points $\mathbf{q}_{\sigma}$ satisfying $ \Re[-\Phi(\mathbf{q}_{\sigma})] \geq \max{\Re(-\Phi(q)})$, $ q \in C_{\mathbb{R}} $, it holds that $<C_{\mathbb{R}}|\Pi_{\sigma} > = 0$ and the associated thimbles do not contribute to the path integration. 
On the other hand, 
it holds that if $<C_{\mathbb{R}},\Pi_{\sigma} >=1$ 
the associated thimbles contribute with the relative weights proportional to ${e}^{ -\Re(\Phi[q(\tau)] )} $. 

\section{Algorithmic solution}

To do calculation, for example, the momentum distribution function  $F(p)$ we are going to combine the Monte Carlo method (MC) 
\cite{Metropolis,Hasting} for searching the critical points $q_{\sigma}$ 
and the finite-difference methods for solving equations (\ref{difur}), (\ref{difur1})
with initial conditions 
obtained by MC method in the small $\epsilon$ vicinity of ${\bar{q_{\sigma}}}$. 
In this section we will be a little bit sloppy in our notation as the 
symbol $\overline{\mathbf{q}}$  can denote a complex valued $(2K-1)$-dimensional variables.

Used here MC method is based on the Metropolis -- Hastings algorithm  \cite{Metropolis,Hasting}, which resides in designing 
a Markov process (by constructing transition probabilities 
$P(\overline{\mathbf{q}}\rightarrow \overline{\mathbf{q}}')$ ), such that its stationary distribution 
to be equal to $w(\overline{\mathbf{q}})$. The derivation of the algorithm starts with the condition of detailed 
balance: 
\begin{eqnarray}
	w(\overline{\mathbf{q}})P(\overline{\mathbf{q}}\rightarrow \overline{\mathbf{q}}') = 
	w(\overline{\mathbf{q}}')P(\overline{\mathbf{q}}'\rightarrow \overline{\mathbf{q}})
\end{eqnarray}
which can be rewritten as 
\begin{eqnarray}
	\frac{P(\overline{\mathbf{q}}\rightarrow \overline{\mathbf{q}}')}{P(\overline{\mathbf{q}}'\rightarrow \overline{\mathbf{q}})} = \frac{w(\overline{\mathbf{q}}')}{w(\overline{\mathbf{q}})}. 
\end{eqnarray}
To increase efficiency of the numerical procedure we are going to separate the transition in two sub-steps:   
the proposal and the acceptance-rejection. 
The transition probability can be written as the product: 
$P(\overline{\mathbf{q}}\rightarrow \overline{\mathbf{q}}') = g(\overline{\mathbf{q}}\rightarrow \overline{\mathbf{q}}') A(\overline{\mathbf{q}}\rightarrow \overline{\mathbf{q}}')$. 
The proposal distribution 
$\displaystyle g(\overline{\mathbf{q}}\rightarrow \overline{\mathbf{q}}')$ is 
the conditional probability of proposing a state $\overline{\mathbf{q}}'$ for given $\overline{\mathbf{q}}$. 
The acceptance distribution $A(\overline{\mathbf{q}}\rightarrow \overline{\mathbf{q}}')$ is 
the conditional probability to accept the proposed state $\overline{\mathbf{q}}'$. 

Inserting this relation in the previous equation, we have 
\begin{eqnarray}
	\frac{A(\overline{\mathbf{q}}\rightarrow \overline{\mathbf{q}}')}{A(\overline{\mathbf{q}}'\rightarrow \overline{\mathbf{q}})} = \frac{w(\overline{\mathbf{q}}')}{w(\overline{\mathbf{q}})}\frac{g(\overline{\mathbf{q}}'\rightarrow \overline{\mathbf{q}})}{g(\overline{\mathbf{q}}\rightarrow \overline{\mathbf{q}}')}.
\end{eqnarray}.

Then it is necessary to choose an acceptance that fulfills detailed balance. 
One common choice is the Metropolis's suggestion:
\begin{eqnarray}
	A(\overline{\mathbf{q}}\rightarrow \overline{\mathbf{q}}') = \min\left(1,\frac{w(\overline{\mathbf{q}}')}{w(\overline{\mathbf{q}})}\frac{g(\overline{\mathbf{q}}'\rightarrow \overline{\mathbf{q}})}{g(\overline{\mathbf{q}}\rightarrow \overline{\mathbf{q}}')}\right).
\end{eqnarray}
This means that we always accept when the acceptance is bigger than 1  
and we can accept or reject when the acceptance is smaller than 1. 
 
 To optimize the MC search of the critical point 
 $\partial \bar{\Phi}[\bar{q}]/\partial\bar{q}|_{\bar{q}=\bar{q}_{k,\sigma}} =0$ 
 we define the  probability  $w(\overline{\mathbf{q}})$ as 
 $w(\overline{\mathbf{q}})=\exp (- b |\partial \bar{\Phi}[\overline{\mathbf{q}}]/\partial\bar{\overline{\mathbf{q}}}|^2)$ 
 with parameter $b \ge 1$. We are free in choosing probability 
 $g(\overline{\mathbf{q}}\rightarrow \overline{\mathbf{q}}')$ as it affect only the 
 efficiency of sampling the main contribution to the integrals 
 and does not change the final result of calculations. 
 For optimization of the MC finding the main contribution to the integral (\ref{FinFp}) the choice 
 of the $g(\overline{\mathbf{q}}\rightarrow \overline{\mathbf{q}}')$ may be the following 
 $g(\overline{\mathbf{q}}\rightarrow \overline{\mathbf{q}}')=
 {e}^{ - \beta \Re(\Phi[\overline{\mathbf{q}}'])}/{e}^{ - \beta \Re(\Phi[\overline{\mathbf{q}}]) }$ 
 with free appropriate fit parameter $\beta$.  
 
The Metropolis -- Hastings algorithm consists in the following steps:\\
\begin{enumerate}
\item{Initialization: pick an initial state point $\overline{\mathbf{q}}$ at random.}
\item{Randomly pick a state $\overline{\mathbf{q'}}$, according to probability 
$\displaystyle g(\overline{\mathbf{q}}\rightarrow \overline{\mathbf{q}}')$.
\item{Accept the state according to the probability $\displaystyle A(\overline{\mathbf{q}}\rightarrow \overline{\mathbf{q}}')$.} 
If not accepted, that means that $\overline{\mathbf{q}}' = \overline{\mathbf{q}}$, and so there is no need to update anything. 
Else, the system transits to $\overline{\mathbf{q}}'$.}
\item{Go to 2 until many $M$ states were generated to "forget" initial $\overline{\mathbf{q}}$ and to obtain 
the average position of the critical point $\langle \overline{\mathbf{q}}\rangle$.}
\item{If $w(\langle \overline{\mathbf{q}}\rangle)\ge0.95$ 
carry out several iterations by complex-valued extension of the Newton's method to produce better approximations 
to the roots (or zeroes) of the complex-valued function 
$\partial \bar{\Phi}[\bar{q}]/\partial\bar{q}|_{\bar{q}=\bar{q}_{\sigma}} =0$. Else go to 2.}
\item{Save the state $\bar{q}_{\sigma}=\langle \overline{\mathbf{q}}\rangle$.}
\item{At random pick an initial point $\overline{\delta \mathbf{q}}$ at  the small vicinity of the zero point of the complex space
 (may be in any given quarter).}
\item{Analogously randomly pick a new state  $\overline{\delta \mathbf{q'}}$ 
 according to the probability 
$\displaystyle g(\overline{\mathbf{q}}_{\sigma}+ \delta \mathbf{q}  \rightarrow   \overline{\mathbf{q}}_{\sigma}+ \delta \mathbf{q'})$.}
\item{Go to 8 many times $M'$ to "forget" initial $\overline{\mathbf{\delta q}}$ state and to obtain 
the average value of the  $\langle \overline{\mathbf{\delta q' }}\rangle$.}
\item{Solve equation (\ref{difur1}) with initial conditions of $\overline{\mathbf{q}}_{\sigma}+ 
\langle \delta \mathbf{q'}\rangle $ to check if $n_{\sigma}=<C_{\mathbb{R}},\Pi_{\sigma}>=1$. 
If $n_{\sigma}=1$ go to 11, otherwise go to 2. 
} 
\item{Solve equation (\ref{difur}) with initial conditions of $\overline{\mathbf{q}}_{\sigma}+ 
\langle \delta \mathbf{q'}\rangle $ until modulus of the integrand in (\ref{FinFp}) will be
smaller of several order of magnitude of its initial value and 
calculate the integral sum related to (\ref{FinFp}).}
\item{Go to 2 many times and calculate the integrals (\ref{FinFp}).}
\end{enumerate}

It is important to notice that it is not clear, in a general problem, which distribution 
$\displaystyle g(\overline{\mathbf{q}}\rightarrow \overline{\mathbf{q}}')$ one should use. 
It is a free parameter of the method which 
has to be adjusted to the particular problem 'in hand'. 
This is usually done by calculating the acceptance rate, which is the fraction of 
proposed samples that is accepted during the last $\displaystyle N$ samples. 
As has been shown theoretically the ideal acceptance rate 
have to be in interval of $23$ -- $50$ \%.

\section{Results of numerical test calculations}
 
Before calculation of the Fourier transform defining the Wigner function (\ref{eq_FP_wignerfunction})  
it is necessary to test the proposed algorithm by calculation of the more simple contour integrals with know 
analytical or numerical answer. We started from consideration of the strongly oscillating and low dimensional integrals, 
which can be typically treated very effectively with the saddle point integration method.  
Presented test calculations provide an interesting benchmark for more realistic path integrals. 
It is also non-trivial from the point of view of a Monte Carlo integration. 

\subsection{The Airy function}

\begin{figure}[htb]
\includegraphics[width=4.8cm,clip=true]{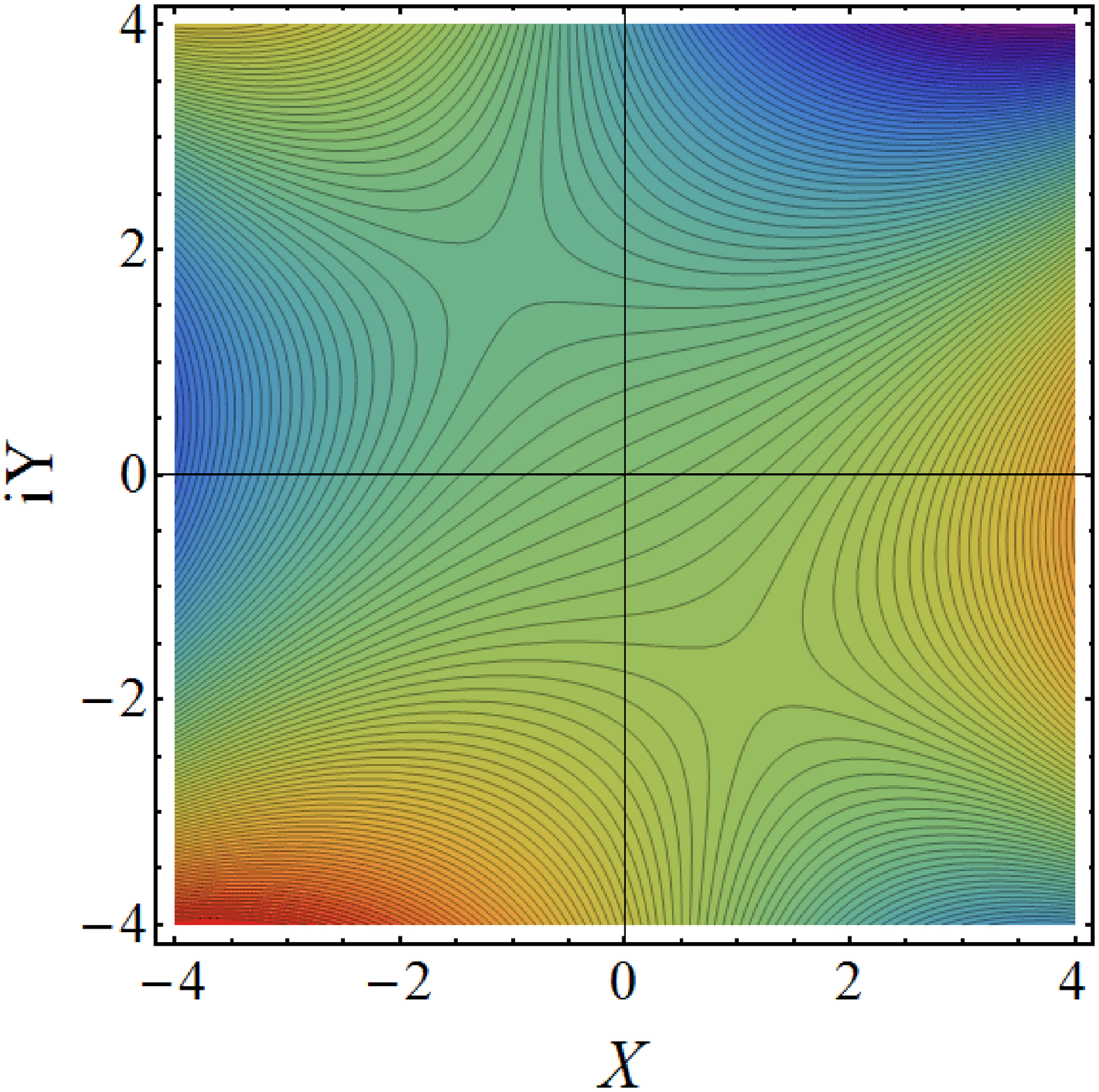}
\hspace {-0.02cm}\includegraphics[width=4.8cm,clip=true]{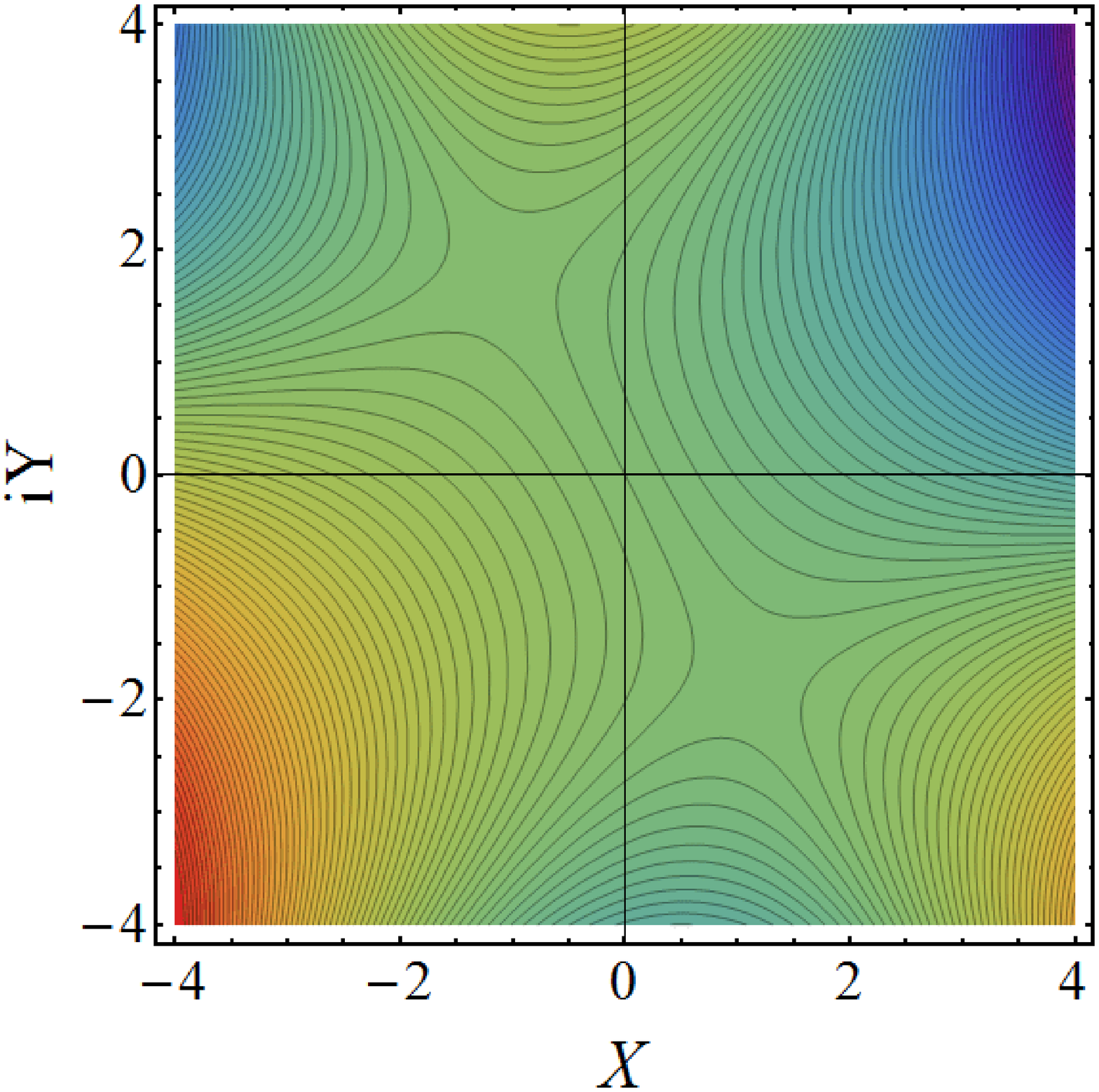} 
\hspace {-0.02cm}\includegraphics[width=4.8cm,clip=true]{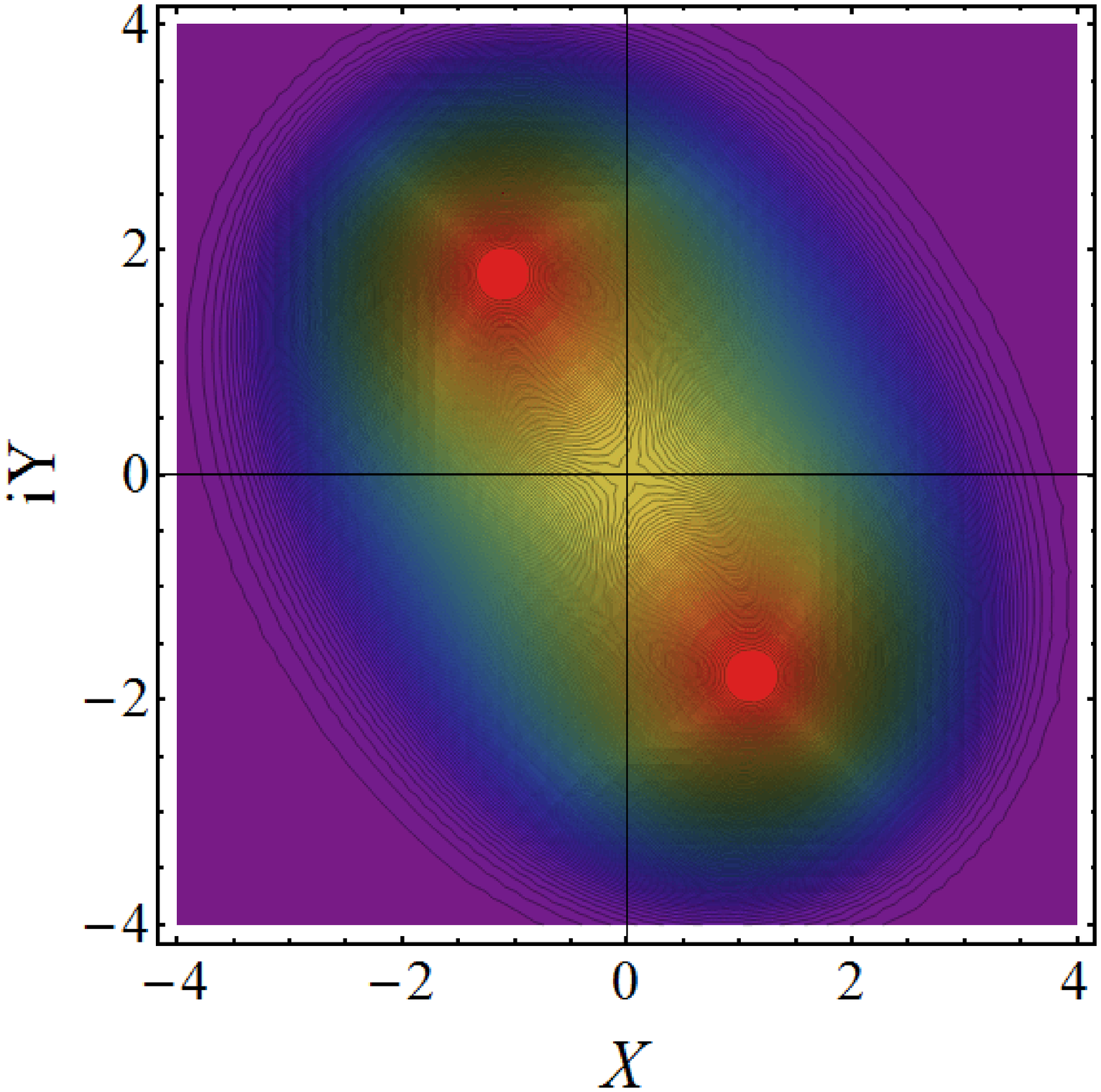}
	\caption{(Color online) 
The contour plot of the imaginary part  (left panel)  and the real part (central panel) of the power in exponent in (\ref{Airy}) 
on complex plane for $p=2+4{\rm{i}}$. (Right plot) Contour plot of the probability 
$w(\overline{\mathbf{q}})=\exp (- b |\partial \bar{\Phi}[\overline{\mathbf{q}}]/\partial\bar{\overline{\mathbf{q}}}|^2)$. 
} 
	\label{cmpxpl}   
\end{figure}

Let us consider a classic example: the Airy function is defined as the integral over the real axis $C_{\mathbb{R}}$:  
\begin{equation}
	\label{Airy}
\rm{Ai}(p)= \frac{1}{2 \pi}\int\limits_{-\infty}^{+\infty} {\rm{d}}x \, \exp \left[{\rm{i}}\left(p x + \frac{x^3}{3} \right) \right],
\end{equation} 
The integrand is strongly oscillating function on $C_{\mathbb{R}}$, which makes a direct numerical evaluation infeasible. 
The left plot of the Fig.~\ref{cmpxpl} shows lines of the constant imaginary part of the power in exponent in (\ref{Airy}) on 
complex plane, while increasing and lowering values are denoted by the red and blue regions. 
The critical points are in the left upper and the right bottom quarters of the complex plane.  
We can deform the integration path in the complex plane of variable $z = x + {\rm{i}} y$, as long as the new path belongs to the original 
relative homology class, which connects regions of strongly decaying modulus of the integrand at infinity 
(blue regions on the centre plot of the Fig.~\ref{cmpxpl}). Right panel of this figure shows the contour plot 
of the MC probability 
$w(\overline{\mathbf{q}})=\exp (- b |\partial \bar{\Phi}[\overline{\mathbf{q}}]/\partial\bar{\overline{\mathbf{q}}}|^2)$ 
with two red circles at its maximum values at the critical points. 


\begin{figure}[htb]
	\includegraphics[width=8.5cm,clip=true]{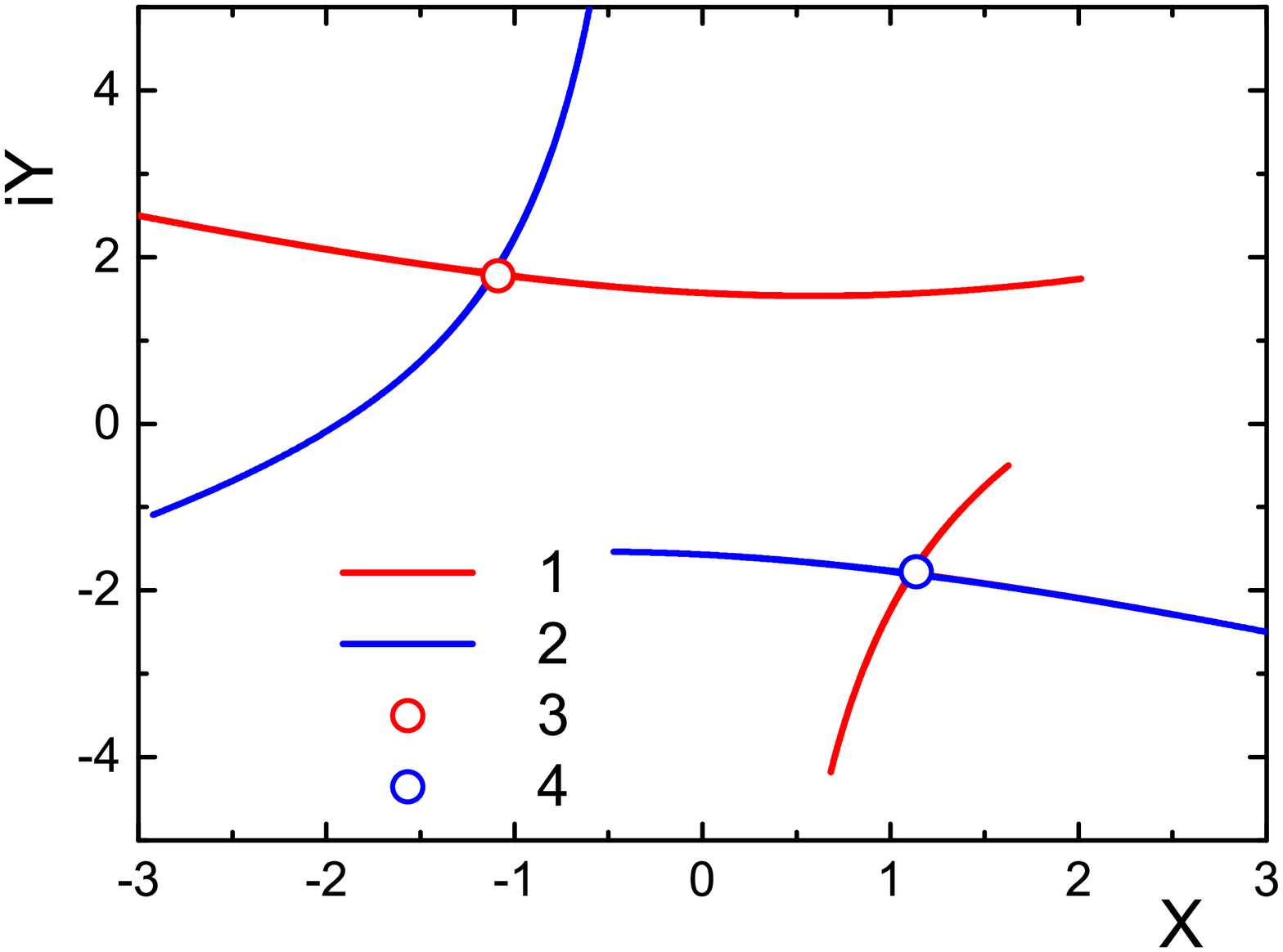}	 
	\includegraphics[width=8.5cm,clip=true]{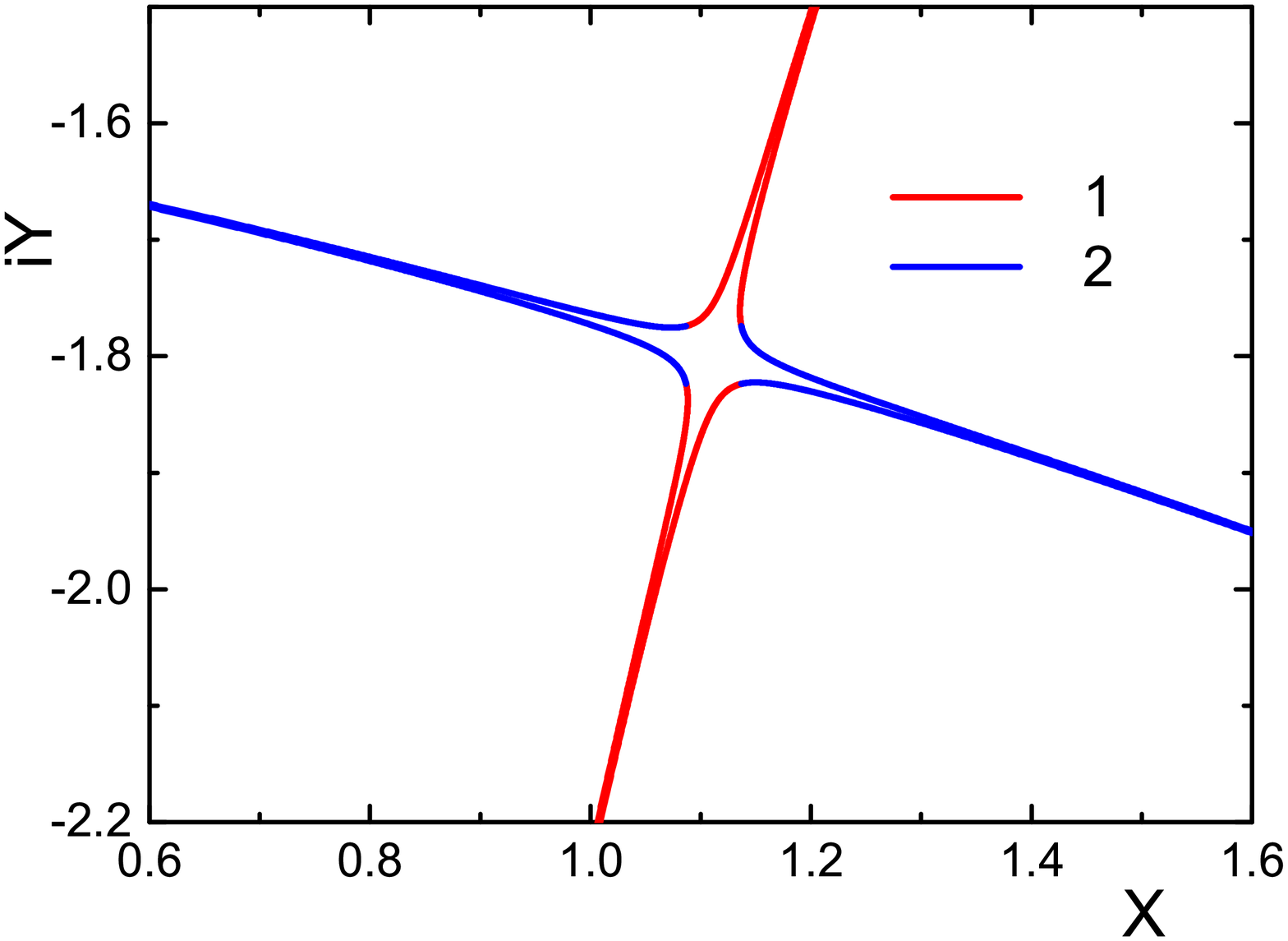}	
	\caption{(Color online) (Left plot)
The averaged downward flows are lines -- 1 ($\in \Upsilon_{\sigma}$)  and the upward flows are 
lines -- 2 ($\in \Pi_{\sigma}$). The critical points are points 3 and 4  
($\partial \bar{\Phi}[\bar{q}]/\partial\bar{q} =0$) of the integrand 
in (\ref{Airy}). Red critical point and the associated thimbles contribute to the contour integral of the Airy function 
(\ref{Airy})  as $<C_{\mathbb{R}},\Pi_{\sigma} >=1$, while the blue point not, as  $<C_{\mathbb{R}},\Pi_{\sigma} >=0$.
(Right plot) Details of initial conditions obtained by MC method in different quarters of small vicinity of the critical 
point. 	}		
	\label{cmpxpll}   
\end{figure}

Testing the proposed approach starts from finding critical points by suggested MC method. 
It turns out that the Markovian chain traveling on the whole complex plane always  stabilizes in the vicinity of the  
the critical point $z_\sigma=-1.11+\rm{i}1.79$ (point in the left upper quarter of the 
complex plane in Fig.~\ref{cmpxpl}) ignoring the second critical point ($z_\sigma=+1.11-\rm{i}1.79$) 
in the favoure of the the first one.  So to force the Markovian chain to stabilize in the vicinity of the second critical point 
we have to use the special restrictions.   
Reason of this behavior of the Markovian chain is the asymmetry of the contour plots of the real and imaginary parts of 
the power in exponent in (\ref{Airy}) (see the Fig.~\ref{cmpxpl} ). 

Solution of the complex valued differential equations (\ref{difur} and \ref{difur1} ) with MC 
initial conditions nearby the both critical points allow to obtain averaged downward  (red lines) and upward 
(blue lines) flows (see both plots of the Fig.~\ref{cmpxpll}). Right plot of this figure shows  in details 
the downward and upward flows from different quarters of the small enough vicinities of the critical points.    
Let us note that initial conditions for red downward and blue upward flows were the same to test their 
fast converge to the related $\Upsilon_{\sigma}$ and $\Pi_{\sigma}$ respectively. 
Let us note that the power low grows on the complex plane of the right part of differential equations  
(\ref{difur}) and (\ref{difur1}) results in 
limitations on the 'time'  $l$ of obtained solutions at needed given accuracy.  

\begin{table}[htb]\label{tab1}
\hspace {0cm}
\caption{The MC results versus the exact Airy function  }
\hspace {3.2cm}\begin{tabular}{|r|c|c|c|c|c|c|}
\hline		
$ \ p \ \  $ & $MC$ & $the Airy function$ \\
\hline
2+\rm{i}4   & $0.3365 - \rm{i}0.065451$ &  $0.3301 - \rm{i} 0.088 $ \\
\hline		
0+\rm{i}4  & $-4.8569 + \rm{i}7.244$ &  $-4.6362 + \rm{i} 7.4111  $ \\
\hline		
4+\rm{i}0  & $0.000738 +\rm{i}0.00006$ &  $0.000952 + \rm{i} 0 $ \\
\hline					
\end{tabular}
\end{table}

As only the blue line for red point crosses the real axis ($<C_{\mathbb{R}}|\Pi_{\sigma} >=1$)  we calculate 
integral defined the Airy function along the red Lefschetz thimble. As we mentioned above the Markovian chain traveling 
on the whole complex plane prefer namely this critical point ignoring the second one. The reason of this interesting 
fact has be further investigated. 

Result of calculation at $p=2+\rm{i}4$ is presented in 
Table 1 as well as results of the some additional calculations for $p=0+\rm{i}4$ and $p=4+\rm{i}0$. 
Comparison of obtained results 
with well known values of Airy function demonstrate a good enough  agreement.  
Discrepancy between exact values of the Airy function and related values obtained by proposed  
procedure can be explained by approximations used in transitions from initial integral 
to its Lefschetz thimbles representation accounting for only the main contribution to the contour integrals.
\subsection{Short time path integral}

Now to test the developed approach let us consider elementary factor (\ref{eq_AS_intikreal}) in the 
finite dimensional approximation of the path integral.  
Expression (\ref{eq_AS_intikreal})  may be rewritten in the form like: 
\begin{eqnarray}
\label{eq_fn_intikreal1}	
\label{fact}
I_{k}(p_k) &=& \int\limits_{-\infty}^{+\infty} {d}q_{k} \, \exp\left\{ \rm{i} (p_kq_k  + \rm{i} \left((q_{k}-const)^2 + q_{k}^4/4 )\right) \right\}  \, ,
\end{eqnarray}
where, for example, $const=2$. 
\begin{figure}[htb]
	\label{eq_fn_intikreal3}	
	\includegraphics[width=4.8cm,clip=true]{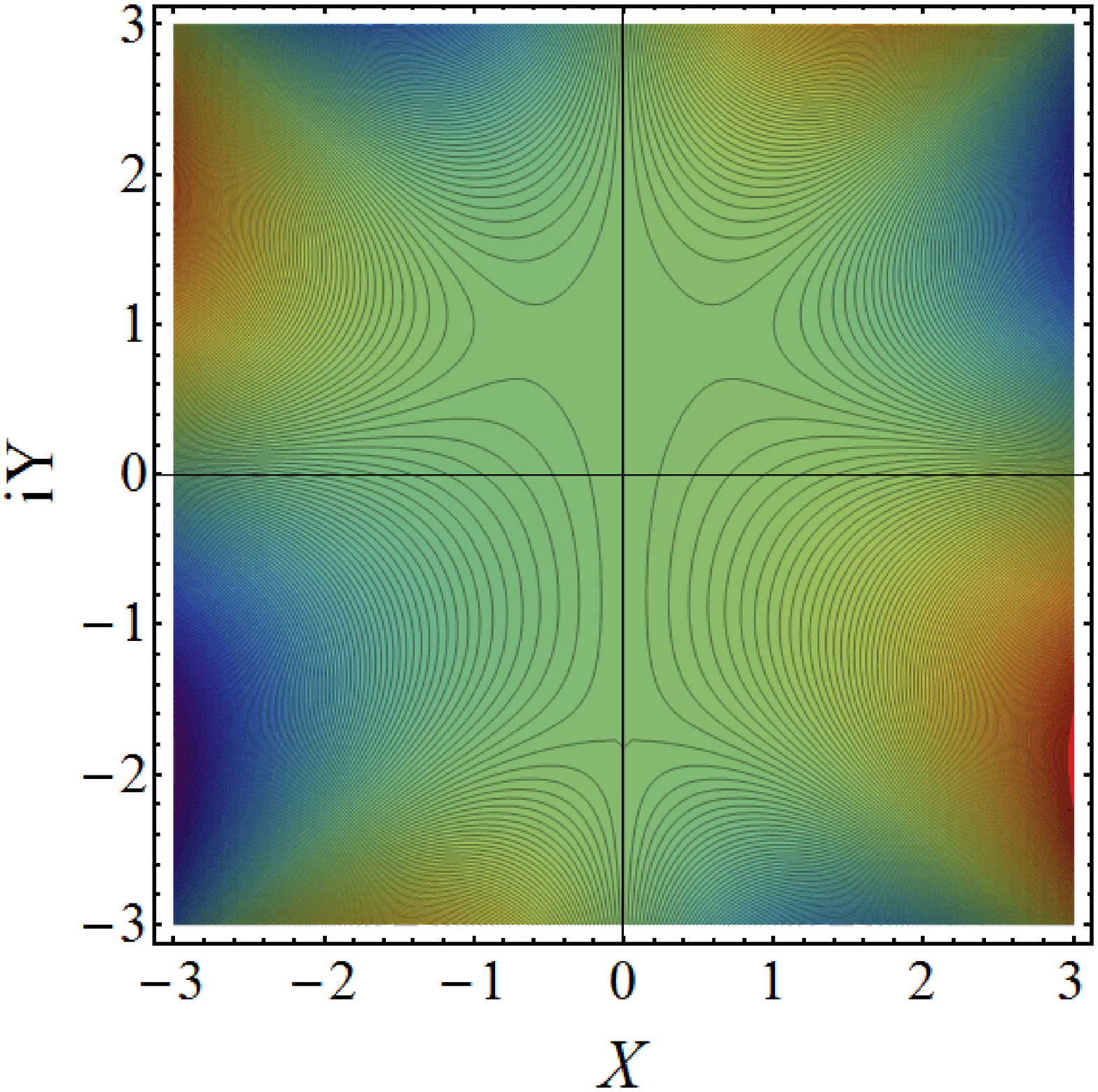}
\hspace {-0.02cm}\includegraphics[width=4.8cm,clip=true]{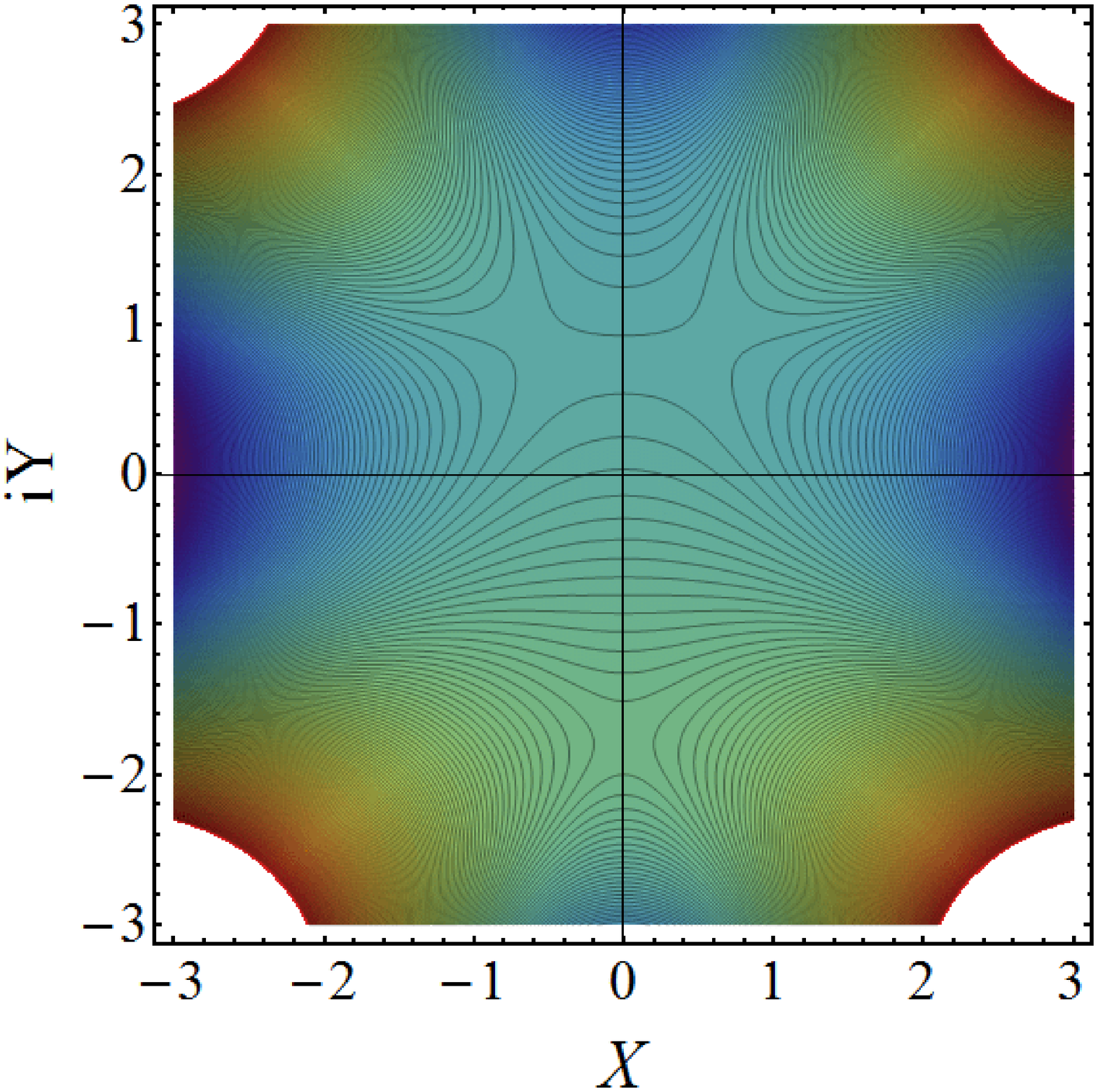}	
\hspace {-0.02cm}\includegraphics[width=4.8cm,clip=true]{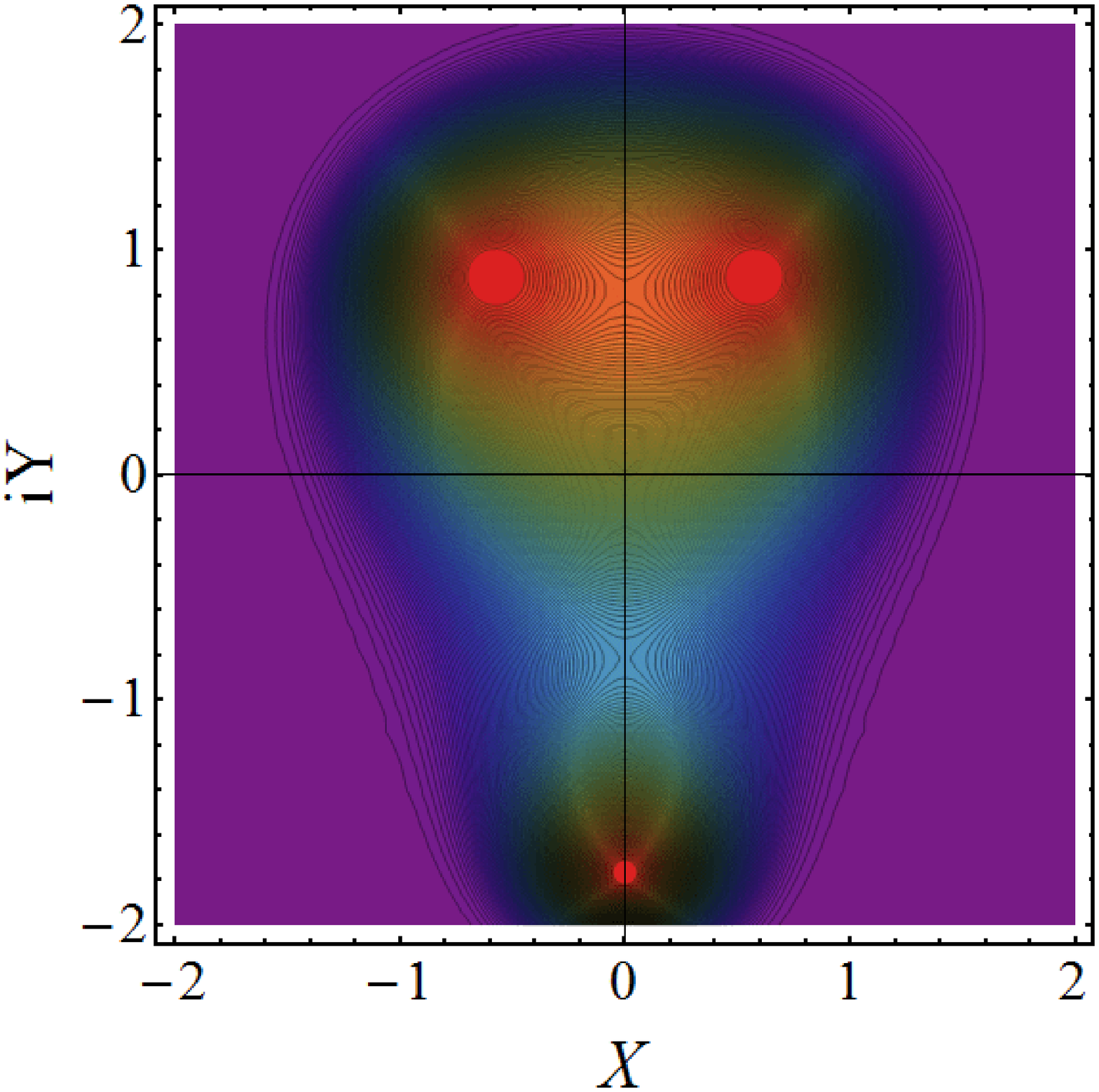}  
	\caption{(Color online) 
The contour plot of imaginary part(left panel)  and real part  (central panel) of the of the power in exponent in (\ref{fact}) 
on complex plane for $const=2$.  (Right plot) Contour plot of the probability 
$w(\overline{\mathbf{q}})=\exp (- b |\partial \bar{\Phi}[\overline{\mathbf{q}}]/\partial\bar{\overline{\mathbf{q}}}|^2)$. 		
} 
	\label{fncmpxpl2}
\end{figure}

The left and right plots of Fig.~\ref{fncmpxpl2} are the contour plots of the image and real parts of the power of exponent 
in integrand in (\ref{eq_fn_intikreal1}) respectively. 
Right plot of the Fig.~\ref{fncmpxpl2} presents contour plot of the probability, 
$w(\overline{\mathbf{q}})=\exp (- b |\partial \bar{\Phi}[\overline{\mathbf{q}}]/\partial\bar{\overline{\mathbf{q}}}|^2)$ 
which with conditional probability $g(\overline{\mathbf{q}}\rightarrow \overline{\mathbf{q}}')$ 
allows to find by MC method the critical points and provides the initial conditions for solution of the equations 
(\ref{difur}) and (\ref{difur1}). 
According to the Monte Carlo simulation the critical points 
are  $z_{1} \approx 0.5712 + \rm{i} 0.8786$, $z_{2} \approx -0.5712 + \rm{i} 0.8786$ and $z_{3} \approx -1.77 + \rm{i} 0.0 $, which
agree with the regions of the saddle-like behaviour of the contour lines on left and right plots of Fig.~\ref{fncmpxpl2}. 
Here the Markovian chain traveling on the whole complex plane always  stabilizes in the vicinity of the  
the critical point $z_1$ and $z_2$ ignoring the point $z_3$.  
\begin{figure}[htb]	
	\includegraphics[width=8.1cm,clip=true]{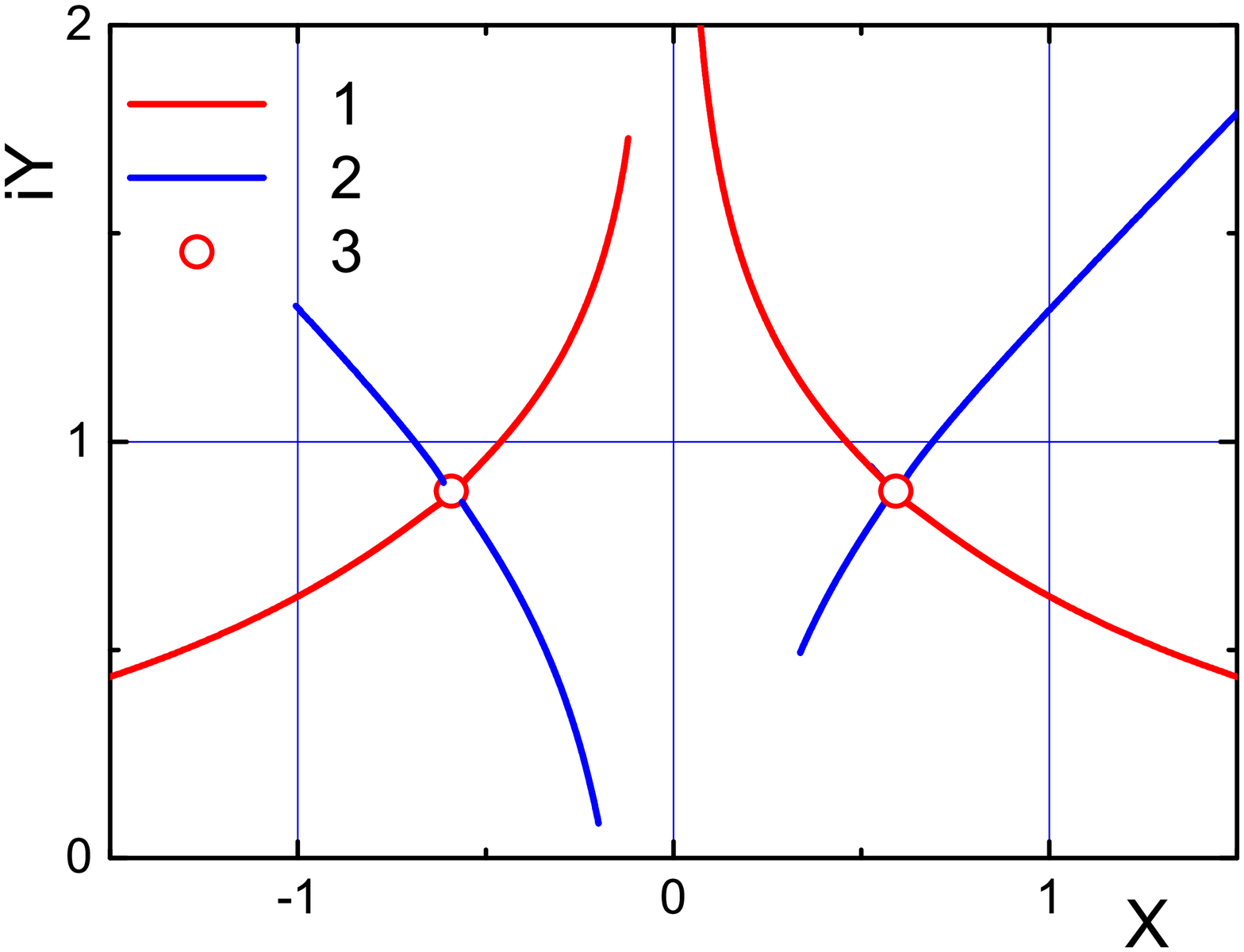}
	\includegraphics[width=8.1cm,clip=true]{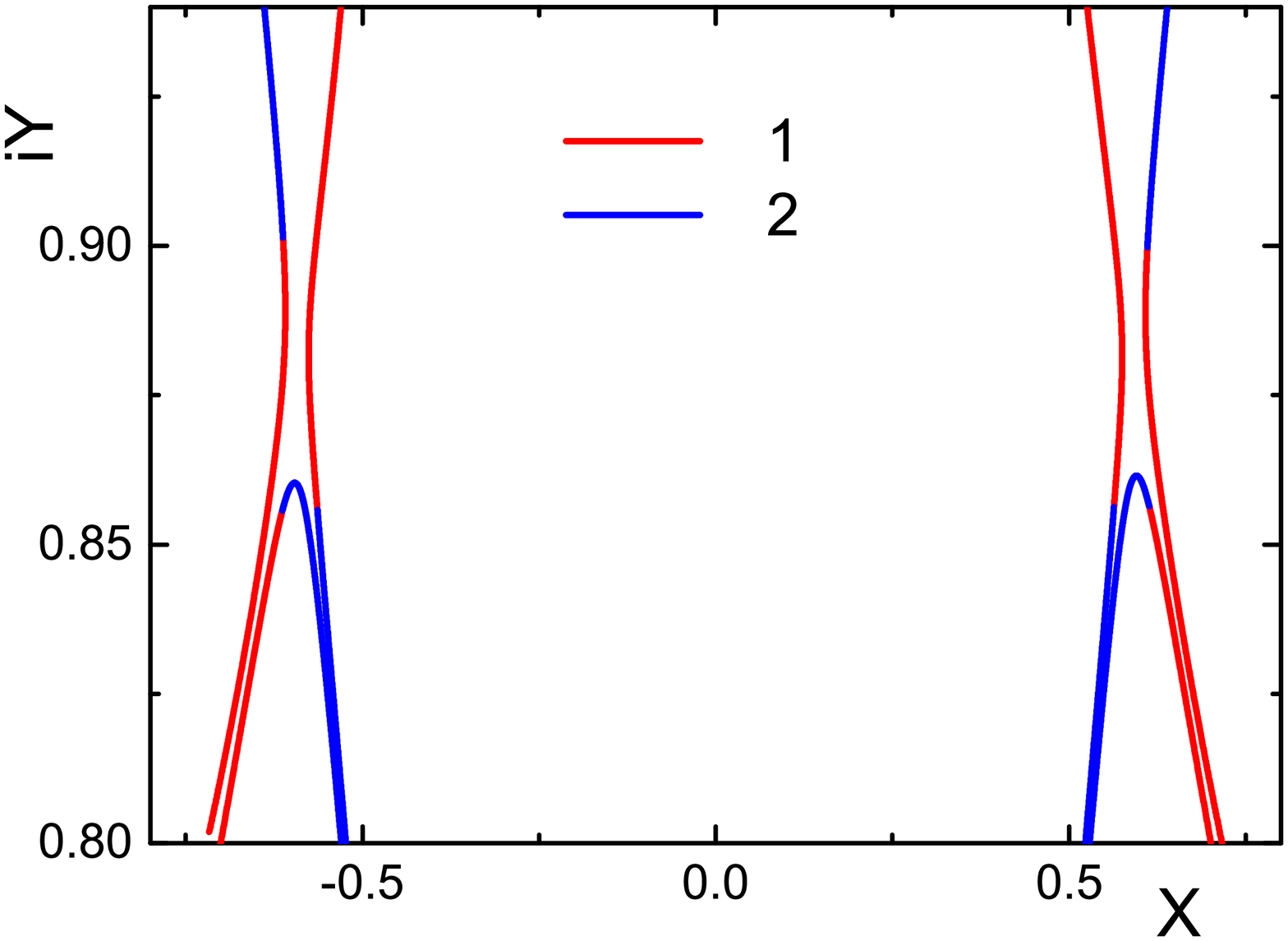} 			
	\caption{(Color online) 
		The averaged downward flows are lines -- 1 ($\in \Upsilon_{\sigma}$)  and the upward flows are 
		lines -- 2 ($ \in \Pi_{\sigma}$) at $const=2$ and $const=2$ and $p=2+\rm{i} 4$ . The critical points 
		of the integrand in (\ref{eq_fn_intikreal1}) are points 3 
		($\partial \bar{\Phi}[\bar{q}]/\partial\bar{q} =0$). Red critical point and the associated thimbles contribute to the contour integral 
		(\ref{eq_fn_intikreal1}) as $<C_{\mathbb{R}},\Pi_{\sigma} >=1$.
	(Right plot) Details of MC initial conditions in different quarters of small vicinity of the critical points.
	}
	\label{fncmpxpl3}
\end{figure}

As before solution of the complex valued differential equations (\ref{difur}) and (\ref{difur1} ) with MC 
initial conditions nearby the both upper critical points allow to obtain averaged downward  (red lines) and upward 
(blue lines) flows (see Fig.~\ref{fncmpxpl3}).  
As the both blue lines for red point cross the real axis ($<C_{\mathbb{R}}|\Pi_{\sigma} >=1$)  we calculate sum 
of the contributions to the integral (\ref{eq_fn_intikreal1}) along the both red Lefschetz thimbles with opposite sign 
due to the different thimble orientation. According to the Morse theory contribution of the bottom critical point 
has to be ignored as the related blue line does not cross real axis (here not shown) and the maximum of the 
integrand on the real axis is smaller than the value of the integrand at this critical point.  

As before details the MC initial conditions and related downward and upward flows are presented by right plot of the 
Fig.~\ref{fncmpxpl3}. Let us note that red and blue lines start at the same MC initial points.  
Reason of asymmetry in behavior of the red and blue lines is in asymmetry of the real and 
imaginary parts of the power in exponent. (see the Fig.~\ref{fncmpxpl3} ). 
Let us stress the fast convergence of the upward and downward flows to the related limit lines 
$\Upsilon_{\sigma}$ and $\Pi_{\sigma}$. 
Comparison of the Monte Carlo result for $I_k=0.01306+\rm{i}0.00006$ with independent 
exact calculations $I_k=0.01402+\rm{i}0$ demonstrate a good enough accuracy of the developed approach. 
\section{'The sign problem' of the real valued path integrals.}
The second type of the 'sign problem' discussed in Introduction arises from requirement of 
antisymmetrization of the real valued matrix elements of the density matrix to account for 
Fermi statistical effects.  In this section to avoid this type of 'sign problem' we are going to discuss   
the effective pair pseudopotential in phase space, which  realizes the Pauli blocking of fermions 
due to the repulsion between identical fermions preventing their occupation of the same phase space cell. 
In this approximation the pseudopotential  is used instead of antisymmetrization of matrix elements. 

The antisymmetrized Wigner function of many-particle electron - hole system in canonical ensemble can be defined 
as a Fourier transform \cite{Wgnr,Tatr} of the matrix element of the density operator \cite{QGP1} in the coordinate representation:
\begin{eqnarray}\label{pathint_wignerfunctionint2}
&&W(p,x)=\sum_{\sigma,\acute{\sigma}} 
\int \rm  d\xi\; 
\delta_{\sigma,\acute{\sigma}}\exp (i\langle \xi |p\rangle)
			\langle x+\xi/2,\sigma|e^{-\beta\hat H}|x-\xi/2,\acute{\sigma}\rangle \,
			\nonumber \\&&
			=\frac{C(M)}{Z\left(N_{eh},V,\beta\right)}
			\sum_{\sigma}\sum_{{ {\rm P}_e}{{\rm P}_{h}}} (-1)^{\kappa_{{\rm P}_e}+ \kappa_{{\rm P}_{h}}}
			{\cal S}(\sigma, {{\rm P}_{e i}} \sigma^\prime)\big|_{\sigma'=\sigma}\,
			\nonumber \\&&
			\times \int {\rm d} \xi \,
			\int {\rm d}q^{(1)} \dots {\rm d}q^{(M-1)}\, 
			\exp\Biggl\{-\pi \frac{\langle \xi|{\rm P}_{e h }+E |\xi \rangle}
			{2M } + i\langle \xi |p\rangle
			-\pi \frac{|{\rm P}_{e h } x-x|^2}{M}
			\nonumber\\&&
			-\sum\limits_{m = 0}^{M-1}
			\biggl[\pi | q^{(m)}-q^{(m+1)}|^2 +
			\varepsilon U\biggl(({\rm P}_{e h } x-x)\frac{m}{M}+x + q^{(m)}
			-\frac{ (M-m) \xi}{2M}+\frac{ m {\rm P}_{e h }\xi}{2M} \biggr)
			\biggr]
			\Biggr\} , 
			\nonumber \\
		\end{eqnarray}
where $C(M)$ is constant and the multidimensional vectors $p$ $q$ $\sigma$ present all momenta, coordinates 
and spin variables of particles. 
%
The antisymmetrization for electrons and holes takes into account the Fermi statistics. 
Here in the sum over all permutations ${{\rm P}_e}{{\rm P}_{ h}}$  we have replaced variables of integration $x^{(m)}$ 
for any given permutation ${{\rm P}_e}{{\rm P}_{ h}}$  by relation 
		\begin{eqnarray}
			x^{(m)} = ({\rm P}_{e h }x-x)\frac{m}{M}+x + q^{(m)}
			-\frac{ (M-m) \xi}{2M}+\frac{ m {\rm P}_{e h }\xi}{2M} \,.
			\label{pathint_variableschange}
		\end{eqnarray}
where $q^{(M)}=q^{(0)}$. Details of deriving the Wigner function (\ref{pathint_wignerfunctionint2})  are presented in 
\cite{Feynm,LarkinFilinovCPP,JAMP,Wiener,NormanZamalin,zamalin,EbelForFil,QGP1}.		

As it was shown in \cite{JPA2017} due to the Fermi repulsion of fermions the main contribution to Fermi statistics in (\ref{pathint_wignerfunctionint2}) comes from 
the pair permutations even at temperatures less the Fermi energy. 
This is the physical reason to take into account only identical and pair permutations and neglect the others.

To avoid difficulties at Monte Carlo simulations due to arising the delta-function after the Fourier transform     
the positive Husimi distributions, being a coarse-grained Wigner function, 
can be used  with a Gaussian smoothing for small phase space cells of parameters $\Delta^2_x$ and $\Delta^2_p$  \cite{Tatr}.
The final expression for Wigner function can be written in the form:

\begin{eqnarray}
	\label{pathint_wignerfunctionint5}
	&&W^H(p,q,) \approx\,
	\frac{C(M)}{Z\left(N_{eh},V,\beta\right) }
	\int \rm dq^{(1)} \dots \rm dq^{(M-1)}\,
			\nonumber\\&&
			\times	
	\exp\Biggl\{
	-\sum\limits_{m = 0}^{M-1}
	\biggl[\pi | q^{(m)}-q^{(m+1)}|^2 +
	\varepsilon U(q + q^{(m)}
	) \biggr] \Biggr\}
	\nonumber\\&&
	\times \exp\Biggl\{\frac{M}{4 \pi}
	\Biggr| i p + \frac{\varepsilon}{2}\sum\limits_{m = 0}^{M-1} \frac{ (M-2m) }{M}
	\frac{\partial U(q+q^{(m)})}{\partial x}
	\Biggr|^2\Biggr\}
		\nonumber\\&&
		\times	
	\sum_{\sigma}\exp (-\beta \sum_{l<t}^{N_e} v^{e}_{ lt})\exp (-\beta \sum_{l<t}^{N_h}
	v^{{ h}}_{ lt})
\end{eqnarray}
where the effective phase space pair pseudopotentials, accounting for quantum statistical effects in pair apptoximation, 
look like:
\begin{eqnarray}
	&&v^{a}_{ lt}=-kT \ln\Biggl\{
	1-\delta_{\sigma_{l,a}\sigma_{t,a}}
	\exp \biggl(-\frac{2\pi|q_{l,a}-q_{t,a}|^2 (1-\frac{\tilde{\Delta}^2_{a,x}/\lambda^2_a}{1+\tilde{\Delta}^2_{a,x}/\lambda^2_a})}{\lambda^2_a}\biggr)
	\exp \biggl(-\frac{|(\tilde{p}_{l,a}-\tilde{p}_{t,a})|^2\lambda^2_a}{(2\pi\hbar)^2(\tilde{\Delta}^2_p/\lambda^2_a)}\biggr)
	\Biggr\} 
	\nonumber\\&& 
\tilde{p}_{t,a}=p_{t,a} + \frac{\varepsilon }{2 }\sum\limits_{m = 0}^{M-1} 
\frac{\partial U(q+q^{(m)})}{\partial q_{t,a}}.
\nonumber	 
\end{eqnarray}
Here $a=e,h $, $\Delta^2_{a,x}=\frac{2}{(\pi^2-2)}$ and $z=1/\sqrt{2}$,  
the interaction energy $U$ is the sum of the two-particle quantum Kelbg -- Coulomb pseudopotentials, 
the $\delta_{\sigma,\acute{\sigma}}$ are the Kronecker symbols. 
The phase space pseudopotentials $v^{a}_{ lt}$ allow us to avoid the famous 'fermionic sign problem' and to realize 
the Pauli blocking for electrons/holes with the same spin without antisymmetrization. 
Note also that the expression (\ref{pathint_wignerfunctionint5}) explicitly contains the term, related to the 
classical Maxwell distribution,  modified by terms accounting for influence of interaction 
on the momentum distribution function. 
To extent the region of applicability of obtained phase space pair pseudopotential, 
$\tilde{\Delta}^2_p$ and $\tilde{\Delta}^2_Q $ can be considered as fit functions with values much smaller than unity.
Our test calculations \cite{JPA2017} have shown that the best fit for 
$\tilde{\Delta}^2_p$ can be written in the form
$\tilde{\Delta}^2_p/\lambda^2_a=0.00505+0.056n\lambda_a^3$, while 
$\tilde{\Delta}^2_{a,x}$ and $\tilde{\Delta}^2_Q$ were of order $0.1$. 

\begin{figure}[htb]
	\includegraphics[width=6.25cm,clip=true]{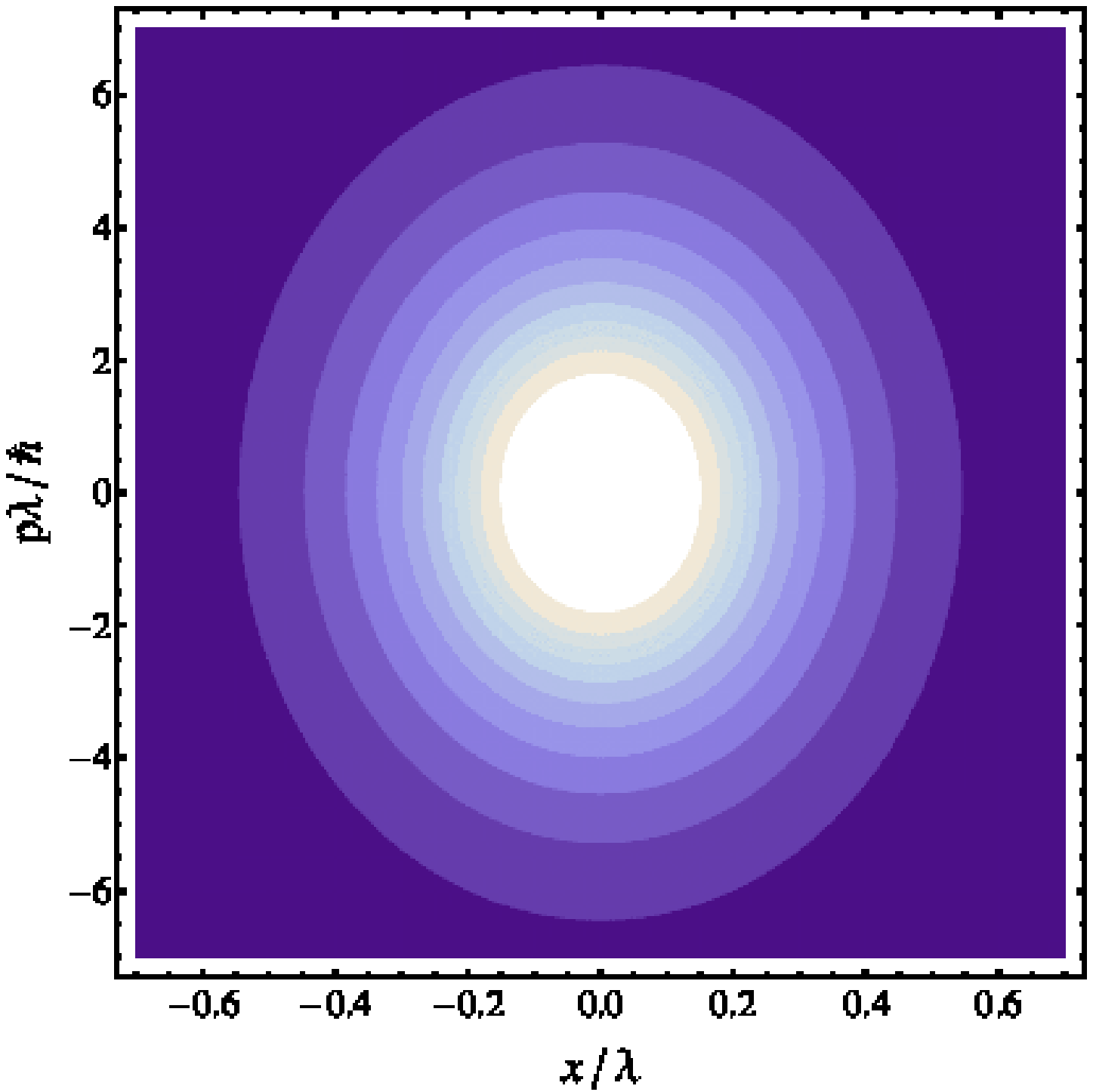}
	\includegraphics[width=10.0cm,clip=true]{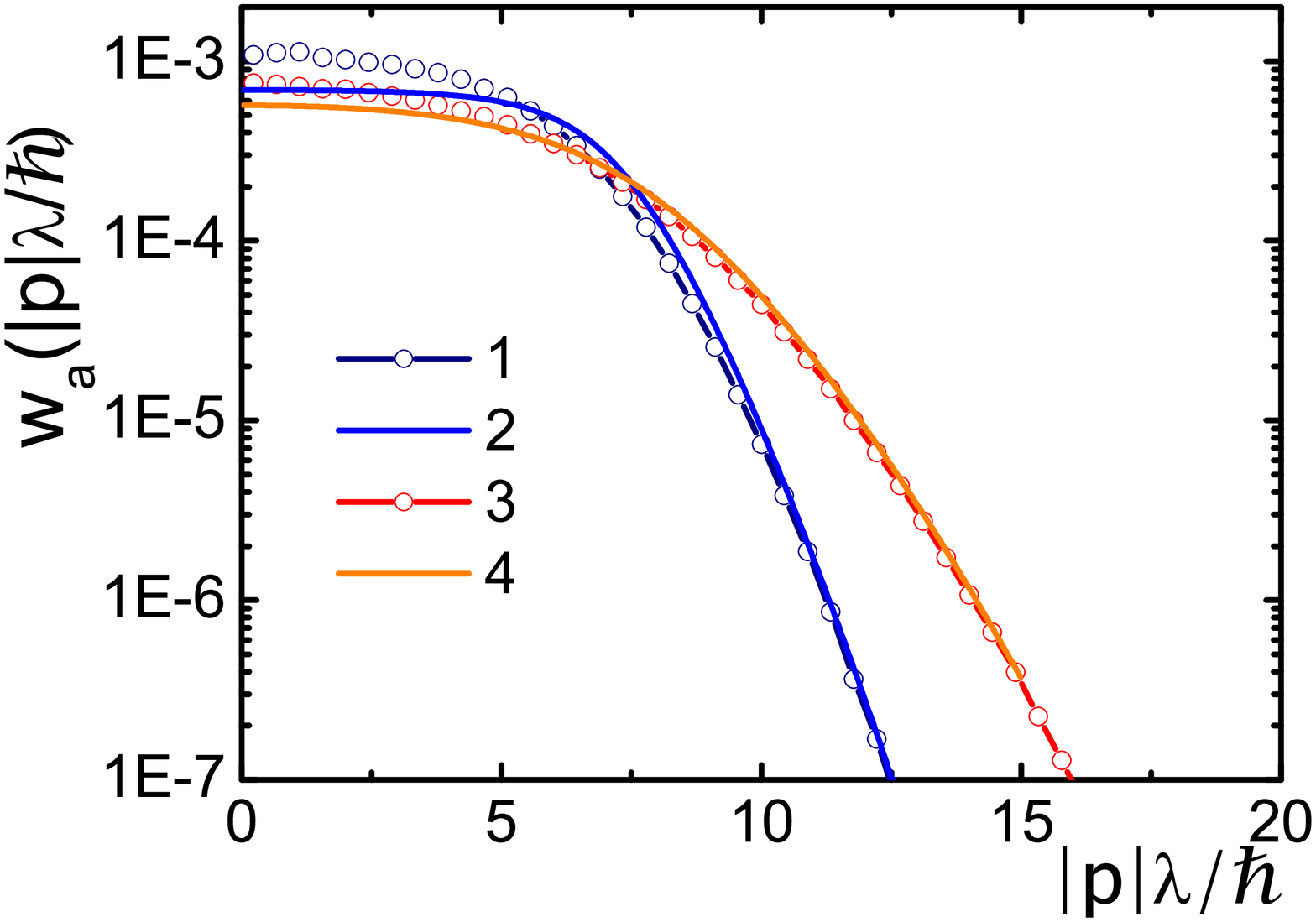}	
	\caption{(Color online)(Left panel) Contour plots of the repulsive effective exchange pair  
pseudopotentials in phase space. Dark area $\beta v^{e}_{ lt}\approx 0$, white area $\beta v^{e}_{ lt}\geq 1.9$. 
(Right panel) The momentum distribution functions $W_a(|p|)$ for ideal electrons (a=e) and two times heavier  holes (a=h).
Lines: $1,\  3$ -- MC distributions $W_a(|p|)$  scaled by ratio of Planck constant to the electron thermal 
(de Broglie) wavelength ($\frac{\hbar}{\lambda_e}$), 
while lines $2, \ 4$ demonstrate the analytical Fermi distributions. 
Parameters of degeneracy $n  \lambda^3_e$ for electrons is 
$n\lambda_e^3=5.6 \, (T/E_F=0.208, \, k_F \lambda_e=5.5) $. 
	}
	\label{ExPt}
\end{figure}

Figure~\ref{ExPt}   presents contour
panels of effective pair  pseudopotentials for parameter of degeneracy $n\lambda_e^3$ 
equal to $5.6 \, (T/E_F=0.208)$. 
Momenta and coordinates  axises are scaled by the electron thermal wave length with Plank constant and factor ten for momentum.

For ideal ($U\equiv 0$ in \ref{pathint_wignerfunctionint5}) electron--hole plasma the Figure~\ref{ExPt} shows
the momentum distributions $W_{a}(|p_a|),(a=e,h)$ 
scaled by ratio of the Planck constant to the electron thermal (de Broglie) wavelength ($\frac{\hbar}{\lambda_e}$). 
In the Figure~\ref{ExPt} results of Monte Carlo calculations for electrons and holes are presented by lines 1 and 3, 
while lines 2 and 4 shows ideal Fermi distributions. Presented distribution functions are normalized to one.
Note that, in these calculations, holes are twice as massive as electrons, so the corresponding degeneracy parameters 
are $2^{3/2}$ times  smaller. It follows from the analysis of Figure~\ref{ExPt} that the agreement between 
the MC calculations and the analytical Fermi distributions  are good enough up to a value of the electron degeneracy parameter 
equal to $n\lambda_e^3 \approx 6$  
where decay of the distribution functions is about of five orders of magnitude.
 
So the Pauli blocking in phase space accounted for by these exchange pseudopotentials
provides agreement of MC calculations and analytical Fermi distribution in wide 
ranges of fermion degeneracy and fermion momenta, 
where decay of the distribution functions is at least of five orders of magnitude.
It necessary to stress that one of the reason of increasing discrepancy with degeneracy growth is limitation on
available computing power allowing calculations with several hundred 
particles  each presented by twenty beads. When parameter of degeneracy is more than 
$10 \, (T/E_F=0.141)$ the thermal 
electron wave length is of order of Monte Carlo cell size and influence of finite number of particles
and periodic boundary conditions becomes significant. 

Let us stress that detailed test calculation for interacting particle carried out within developed approach in 
\cite{LarkinFilinovCPP,JAMP}  have demonstrated good agreement with results of independent calculations 
by standard path integral Monte Carlo method and available analytical results.

\section{Conclusion}

One of the main difficulty for the Path Integral Monte Carli simulation of the quantum systems 
of particles is so called 'sign problem'. However nowadays the term 'sign problem' is used to identify two different 
problems. 

The ideas to overcome the first type of the 'sign problem' of strongly oscillating complex valued integrand in the 
Feynman path integrals comes from Picard-Lefschetz theory and a complex version of Morse theory. 
The main idea is to select Lefschetz thimbles as the cycle 
approaching the saddle point at the path-integration, where the imaginary part of the complex action  
stays constant. Since the imaginary part of the action is constant on each thimble, 
the sign problem disappears and the integral can be calculated much more effectively. 
In this work some simple test calculation and comparison with available analytical results have been carried out. 
We hope that the presented simple testing is a perfect playground to see 
thimble regularization at work. It is instructive both for inspecting 
the structure of relevant thimbles and from the algorithmic point of view.

The second type of the 'sign problem' arises at studies the Fermi systems by path integral approach and is 
caused  by the requirement of antisymmetrization of the real valued matrix elements of the density matrix. 
To overcome this issue the new numerical version of the Wigner approach 
to quantum mechanics for treatment thermodynamic properties of degenerate systems of
fermions has been developed.  Here the main idea is to derived pseudopotential 
in phase space depending on the coordinates and momenta, which due to the repulsion between 
identical fermions prevents their occupation of the same phase space cell and so realizes the Pauli blocking. 
To test the developed approach calculations of the momentum distribution function of the 
degenerate ideal system of Fermi particles has been presented in a good agreement with
analytical Fermi distributions. 
%

\section{Acknowledgements}
We acknowledge stimulating discussions with 
Profs.	Yu.B.~Ivanov. 

\bibliographystyle {apsrev}
\bibliography{FilinovL}

\end{document}